\documentclass{nime-alternate}

\begin{document}  
%
\conferenceinfo{NIME'11,}{30 May--1 June 2011, Oslo, Norway.}

\title{Strike on Stage: a percussion and media performance}

\numberofauthors{2}
\author{
\alignauthor Charles Martin\\
		 \affaddr{Department of Music and Media,}\\
		 \affaddr{Lule{\aa} Technical University}\\
       \affaddr{Pite{\aa}, Sweden}\\
       \email{cpm@charlesmartin.com.au}
\alignauthor Chi-Hsia Lai\\
       \affaddr{Media Lab, Department of Media,}\\
       \affaddr{School of Art and Design, Aalto University}\\
       \affaddr{Helsinki, Finland}\\
       \email{me@laichihsia.com}
}

\maketitle
\begin{abstract}
This paper describes \emph{Strike on Stage}, an interface and corresponding audio-visual performance work developed and performed in 2010 by percussionists and media artists Chi-Hsia Lai and Charles Martin. The concept of \emph{Strike on Stage} is to integrate computer visuals and sound into an improvised percussion performance. A large projection surface is positioned directly behind the performers, while a computer vision system tracks their movements. The setup allows computer visualisation and sonification to be directly responsive and unified with the performers' gestures.
\end{abstract}


\section{Integrating Media Performance}
\begin{figure}
\centering
\includegraphics[width=\columnwidth]{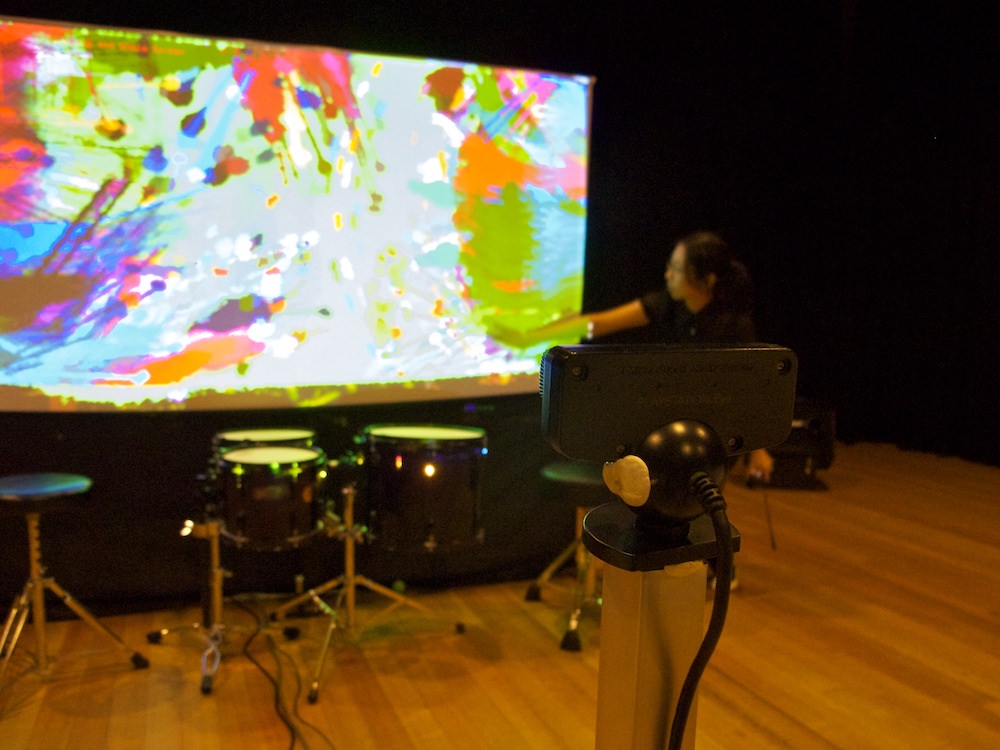}
\caption{\emph{Strike on Stage} in performance.}
\label{Strike on Stage}
\end{figure}

\emph{Strike on Stage} is an interface and corresponding audio-visual performance developed and performed in 2010 by percussionists and media artists Chi-Hsia Lai and Charles Martin. The aim of the work was to integrate computer visuals and sound into an improvised percussion performance. This integration takes place in two ways. First, the performers' gestures are linked to computer audio and visuals through a computer vision system and microphone. Secondly, the presentation of the performance is unified. The projection screen and loudspeakers are placed immediately behind the performers and the computer visuals and audio is designed in such a way that, to the audience, they appear to be a natural augmentation of the performers' forms, instruments and gestures.


The work is a successor to Lai's performance work \emph{Hands on Stage}~\cite{lai-hos-nime2009}, developed at the Australian National University. In \emph{Hands on Stage}, Lai used a computer vision surface similar to that commonly used for \emph{reacTIVision}~\cite{KaltenbrunnerBencinca-reactivision}. In the case of \emph{Hands on Stage}, the focus was not on making a touchable GUI, since there was no projector under the acrylic surface, but rather an extended musical and media instrument. Sounds were produced by the shadow of the player's hands on the surface, detected by a camera, and contact microphones amplified the sound of the player scratching and tapping the surface. \emph{Hands on Stage} also had a visual component influenced by the image of the player's hands which was projected onto an external screen.

\emph{Strike on Stage} was conceived to further the artistic direction of \emph{Hands on Stage} while addressing some of the limitations of the interface. Whereas \emph{Hands on Stage} was designed for a solo performer using only their hands, \emph{Strike on Stage} was designed for two performers, using their whole arms, bodies and drum sticks or other percussion implements to control the performance. The video projection was to be integrated into the performance surface so that the audience's focus is not divided between the performers and an external screen.

\section{The Interface}
\begin{figure}
\centering
\includegraphics[width=\columnwidth]{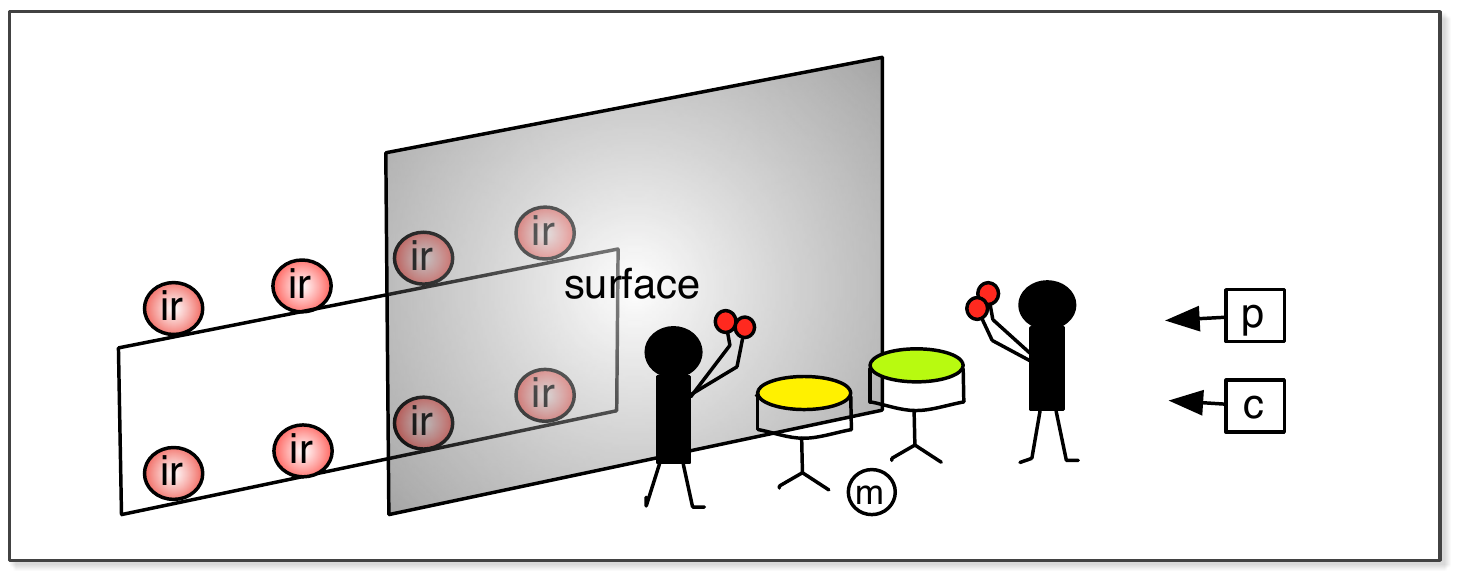}
\caption{Setup diagram for \emph{Strike on Stage}.}
\label{setup}
\end{figure}

The setup for \emph{Strike on Stage} is centred around a floor-standing  projection screen made from thin fabric (denoted `surface' in figure \ref{setup}). An array of 8 infra-red LED security lights (denoted `ir') is placed behind the surface and the lights aimed to provide an even illumination of the screen.

The performers and instruments are positioned directly in front of the screen. Infra-red light passes through the screen from behind enabling an infrared sensitive camera (`c') placed around 3 meters in front of the performers to see a clean silhouette of the performers and instruments. A projector (`p'), outputting mainly visible light is also placed in front of the screen to project onto both the screen and performers. Additionally, a microphone (`m') is placed underneath the instruments and loudspeakers are hidden with all computer equipment behind the screen and IR lights. 

This setup gives a clean image for computer vision purposes and integrates the projection surface with the performers and instruments. Furthermore, the screen and lighting array is lightweight and portable and can be reproduced relatively cheaply for future projects.

Since the computer equipment and loudspeakers are hidden behind the screen, the setup for \emph{Strike on Stage} presents a relatively minimalistic stage presence to the audience. 

\section{The Performance}
\begin{figure}
\centering
\includegraphics[width=\columnwidth]{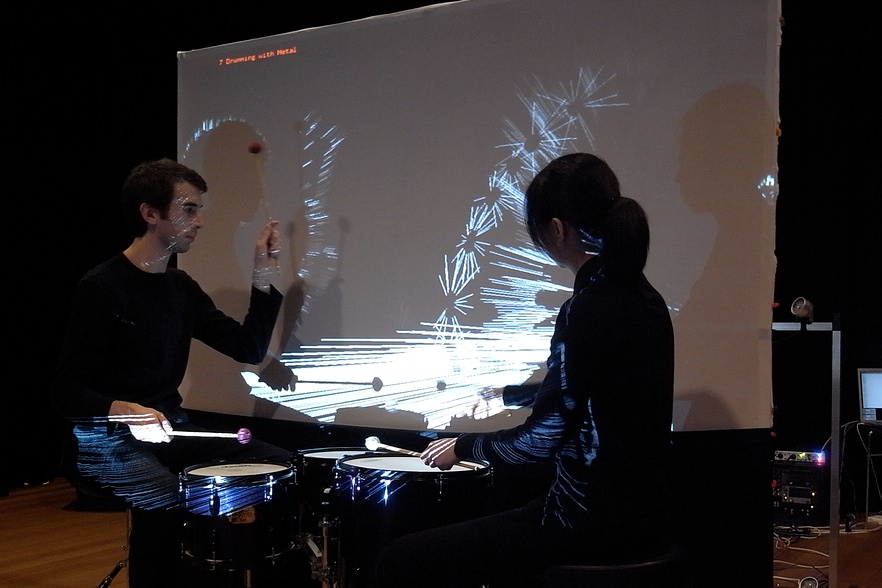}
\caption{One interactive scene in the performance.}
\label{linesScene}
\end{figure}

The performance component of \emph{Strike on Stage} consists of a series of interactive set pieces created in openFrameworks~\cite{openframeworks-website} and SuperCollider~\cite{supercollider-website}, along with percussion improvisations. The work focuses on exploring how the interactive environment can augment the percussive gesture of `striking' an instrument.

Percussion performance is often characterised by the visual drama of striking instruments as much as their sound. As a result, part of a percussionist's individual style is their approach to striking instruments, both in order to produce particular sounds and to emphasise elements of the performance to the audience.

The interactions developed for \emph{Strike on Stage} augment and react to these `percussive gestures'. The technical approach for detecting these movements from blob-tracking algorithms provided in openFrameworks is simple. The performers are positioned side-on to the screen and facing each other (as in figure \ref{setup}). This means that the tip of the left performer's sticks or hands are the rightmost point of their silhouette and similarly for  the leftmost point of the right performer's silhouette. Percussive gestures can be detected by tracking the acceleration of these points. When a drumstick bounces off a drum it has a high acceleration away from the drum at the point of impact. 

In one of the most effective interactions in \emph{Strike on Stage},  manipulated cymbal and gong samples were played in SuperCollider each time a sharp acceleration was detected at the tips of each performer's sticks.  Lines were projected around the edges of the performer's silhouettes that varied their length with the acceleration of that point on the silhouette (shown in figure \ref{linesScene}). The result was that the performers could not only play computer sounds by striking their physical instruments but also by `air drumming'. Since the rationale for triggering extra sounds was simple the performers could control the extent of the effect and play with the interaction in their improvisation. From the audience's perspective, the performers appeared to be surrounded by constantly shifting spines that shot out in synch with the energy of their motions.

Other interactions in the work used the performer's motions to trigger and manipulate field recordings and photographs, both taken by the performers while producing the work. The result was a collage linking the collaborative improvisation with audio and visual textures from the development of the work.

The overall tone of the work was a playful exploration of the affordances of the screen and computer sound. Artistically the work emphasised the `strike' movement of playing percussion instruments and made connections between the performer's movements on stage and their lived experience as creators of the work.

\section{Conclusions and Further Work}
\emph{Strike on Stage} was performed in 2010 as \emph{Strike on Stage 1.0}, and performances were held at Belconnen Arts Centre, Canberra, NIME2010, Sydney and the Australasian Computer Music Conference 2010, Canberra. The work was also converted into a `micro' version with a much smaller screen and only one IR light. These performances are documented on the project's blog\footnote{\url{http://strikeonstage.posterous.com}}.

These performances proved that the setup for \emph{Strike on Stage} was viable in a range of performance conditions even when setup time was extremely limited. Furthermore, feedback from the audience confirmed that the performance method was interesting and effective.

The strategy for capturing `percussive gestures' from blob-tracking algorithms was reasonably effective, but there is much scope to explore other connections between the performers' movements and computer sound and visuals.

Although there are plans to revise \emph{Strike on Stage} with a new version in 2011 the same techniques could inspire other artistic projects. We imagine a collaboration with a composer or an ensemble with multiple `micro' screens and small projectors.

\section{Acknowledgments}
This project was supported by the A.C.T. Government, Australia. 

\bibliographystyle{abbrvurlfornime}
\bibliography{cpm-2010-NIMEpaperBib}  
\end{document}